\documentclass[a4paper,11pt]{article}
\usepackage{pos}
\usepackage{url}
\usepackage[utf8]{inputenc}

\DeclareUnicodeCharacter{0087}{\={A}}

\newcommand{\pe}{p.e.} 

\title{Camera Calibration of the CTA-LST prototype}
 \ShortTitle{Camera Calibration of the CTA-LST prototype}

\manuallySeparateAuthors
\author*[a]{Yukiho Kobayashi}
\author[b]{, Akira Okumura}
\author[c]{, Franca Cassol}
\author[d]{, Hideaki Katagiri}
\author[e]{, Julian Sitarek}
\author[e]{, Paweł Gliwny}
\author[f]{, Seiya Nozaki}
\author[d]{and Yuto Nogami}
\author[]{for the CTA LST project}

\affiliation[a]{Institute for Cosmic Ray Research, The University of Tokyo, Chiba, Japan}
\affiliation[b]{Institute for Space–Earth Environmental Research, and Kobayashi–Maskawa Institute for the
Origin of Particles and the Universe, Nagoya University, Nagoya, Japan}
\affiliation[c]{Aix Marseille Univ, CNRS/IN2P3, CPPM, Marseille, France}
\affiliation[d]{College of Science, Ibaraki University, Ibaraki, Japan}
\affiliation[e]{Faculty of Physics and Applied Informatics, University of Lodz, ul. Pomorska 149/153, 90-236 \L\'od\'z, Poland}
\affiliation[f]{Department of Physics, Kyoto University, 606-8502 Kyoto, Japan}

\emailAdd{yukihok@icrr.u-tokyo.ac.jp}

\abstract{The Cherenkov Telescope Array (CTA) is the next-generation gamma-ray observatory that is expected to reach one order of magnitude better sensitivity than that of current telescope arrays. The Large-Sized Telescopes (LSTs) have an essential role in extending the energy range down to 20 GeV. The prototype LST (LST-1) proposed for CTA was built in La Palma, the northern site of CTA, in 2018. LST-1 is currently in its commissioning phase and moving towards scientific observations. The LST-1 camera consists of 1855 photomultiplier tubes (PMTs) which are sensitive to Cherenkov light. PMT signals are recorded as waveforms sampled at 1 GHz rate with Domino Ring Sampler version 4 (DRS4) chips. Fast sampling is essential to achieve a low energy threshold by minimizing the integration of background light from the night sky. Absolute charge calibration can be performed by the so-called F-factor method, which allows calibration constants to be monitored even during observations. A calibration pipeline of the camera readout has been developed as part of the LST analysis chain. The pipeline performs DRS4 pedestal and timing corrections, as well as the extraction and calibration of charge and time of pulses for subsequent higher-level analysis. The performance of each calibration step is examined, and especially charge and time resolution of the camera readout are evaluated and compared to CTA requirements. We report on the current status of the calibration pipeline, including the performance of each step through to signal reconstruction, and the consistency with Monte Carlo simulations.}

\FullConference{37$^{\rm{th}}$ International Cosmic Ray Conference (ICRC 2021)\\
		July 12th -- 23rd, 2021\\
		Online -- Berlin, Germany}

\begin{document}
\maketitle

\section{Introduction}

The Cherenkov Telescope Array (CTA) will be the next-generation Imaging Atmospheric Cherenkov Telescope array that will be observing gamma rays within 20 GeV - 300 TeV energy range with unprecedented sensitivity. Large-Sized Telescopes (LSTs) are dedicated to observation of the lowest part of the CTA energy range. LST-1, the LST prototype in the northern site of CTA, has been commissioned since its inauguration in 2018, and now it's moving towards scientific observations.

The LST-1 camera is composed of 1855 photomultiplier tubes (PMTs) with high quantum efficiency (QE) \cite{sakurai2019calibration}. PMT signals are sampled at 1 GHz by Domino Ring Sampler version 4 (DRS4) chips and digitized by an analog-to-digital converter (ADC). Fast sampling is essential to achieve the low energy threshold by reducing contamination of night sky background (NSB) light which enters each pixel with a rate of $\sim250$ MHz \cite{Masuda:2015yna}. The camera readout has two channels with different amplification, high gain (HG) and low gain (LG) channels, so that the wide dynamic range from 1 photoelectron (\pe{}) to 3,000 \pe\ is covered. Several calibration tests have been performed for commissioning of the LST-1 camera \cite{camera_commissioning}. 

The calibration chain for LST-1 data has been developed as part of LST analysis pipeline \texttt{cta-lstchain} \cite{lstchain}. The calibration chain performs DRS4 pedestal corrections, absolute charge calibration by the F-factor method and signal reconstruction. In this contribution, we report on each calibration procedure performed by the calibration chain and on the performance of signal reconstruction achieved with it.

\section{DRS4 pedestal corrections}

DRS4 chips have intrinsic pedestal characteristics which should be corrected by analysis for minimizing pedestal noise in readout waveforms \cite{Nozaki:2020jis}. The major characteristics to be corrected are offset of individual capacitors, dependence of offset on $\Delta t$ which is defined as the time since the last reading of the capacitor, and spikes which are jumps of offset values for particular capacitors under certain conditions. Our calibration chain deals with all of these systematic effects. Offset of individual capacitors can be corrected by referring to the average pedestal value of each capacitor which can be obtained from a dedicated pedestal run. The dedicated pedestal run is taken every night to calibrate all the data taken during the same night. The capacitor shows additional offset depending of $\Delta t$. It is known that the $\Delta t$ dependence is well described by a power-law function and thus it can be corrected by analysis \cite{2013NIMPA.723..109S}. Finally, spikes appear for specific capacitors which are determined by the position of the previous readout window in the capacitor array. The spike positions are predictable and thus the spikes can also be corrected. Currently the spikes are removed by interpolation using samples outside the spike position, but correction by just subtracting ADC sample at spike positions is now being implemented.

\begin{figure}
    \centering
    \includegraphics[width=0.49\textwidth]{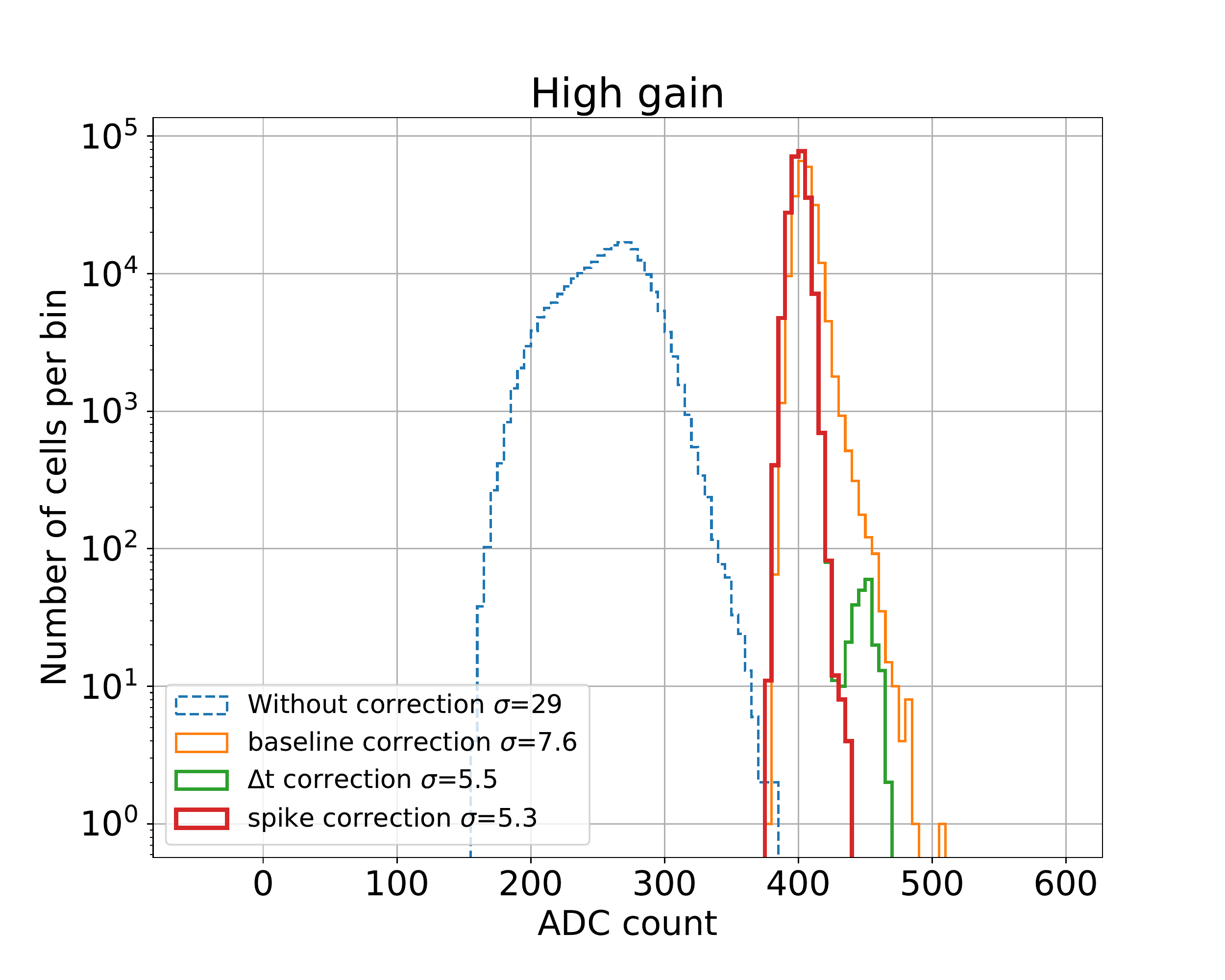}
    \includegraphics[width=0.49\textwidth]{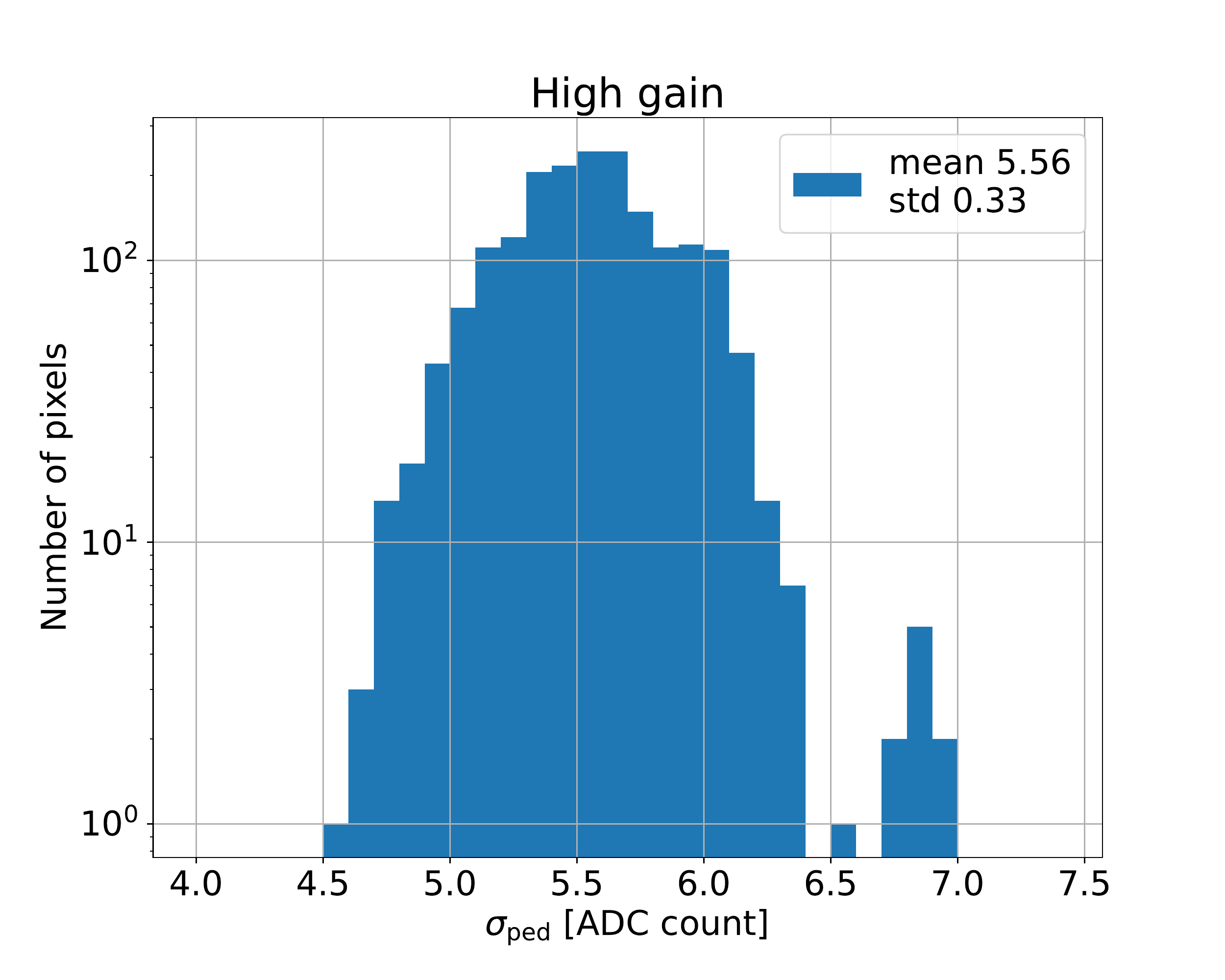}
    \caption{Left: Pedestal distribution in one HG channel after each step of DRS4 pedestal corrections. The value of $\sigma$ is standard deviation of each distribution. Baseline is set to 400 ADC count after the corrections. Right: Distribution of pedestal standard deviation in HG channels after all the corrections. }
    \label{fig:pedestal}
\end{figure}

Fig.~\ref{fig:pedestal} shows the result of pedestal corrections by the calibration chain. The left panel of Fig.~\ref{fig:pedestal} shows pedestal distribution in one channel after each step of the corrections. It can be seen that pedestal noise is reduced by each correction step. Pedestal standard deviation in each HG channel after all the corrections is shown in the right panel of Fig.~\ref{fig:pedestal}. Average pedestal noise after all the corrections is 5.6~ADC count in HG and 3.4 ADC count in LG. This is compatible with ~0.2 \pe\ in HG and ~3 \pe\ in LG.

\section{Absolute charge calibration by the F-factor method} \label{sec:charge_calib}
Absolute charge calibration, i.e., conversion of signal integrated charge in ADC counts to the number of \pe\ produced by light pulses in each PMT, is obtained with the so-called F-factor method. This is based on flat-field events achieved by the uniform illumination of the camera with the diffused light emitted by a laser ($\mathrm{\lambda}$=355 nm) placed in the center of the telescope mirror dish \cite{2019ICRC757}. The number of \pe\ associated with each waveform is estimated by analyzing the first and the second order moments of the charge distribution in a high-intensity regime ($\sim80$ \pe{}/pulse). For each pixel, the median and the standard deviation of the charge are estimated on a sample of flat-field events ($\overline{Q}$ and $\sigma_Q$) and a sample of pedestal events ($\overline{\rm ped}$ and $\sigma_{\rm ped}$) including NSB. Then, in case of pure statistical noise, the number of \pe\ per pixel can be estimated as
\begin{equation}
{\rm \pe} = \frac{(\overline{Q}-\overline{\rm ped})^2}{\sigma^2_Q-\sigma^2_{\rm ped}} \mathrm{F^2},
\label{eq:F-factor}
\end{equation}
where ${\rm F}^2=1\,+\,\sigma_{\rm spe}^2$ is the squared excess noise factor and $\sigma_{\rm spe}$ is the width of the charge distribution produced by single \pe\ in units of \pe\ \cite{BENCHEIKH1992349}. The average value of $\mathrm{F^2}$ for LST-1 PMTs has been evaluated to be 1.222 by measurement at the laboratory. The value is assumed to be equal for all the pixels in the current analysis.

\begin{figure}
    \centering
    \includegraphics[width=0.49\textwidth]{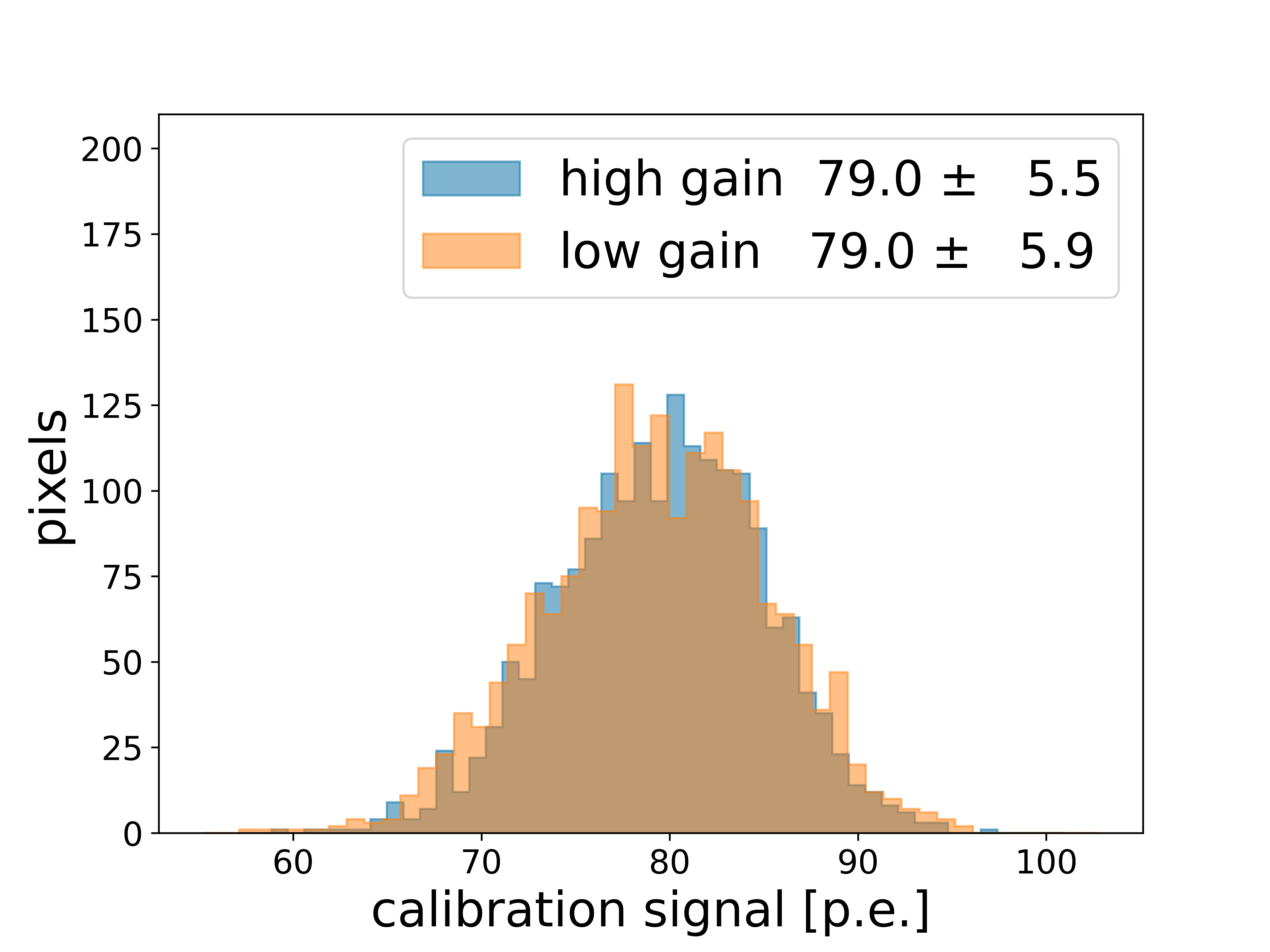}
    \includegraphics[width=0.49\textwidth]{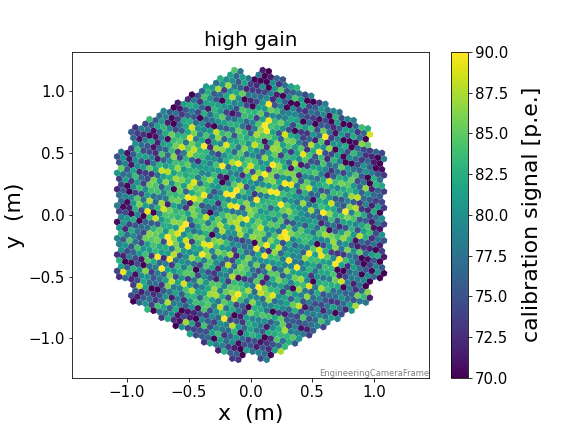}
    \caption{Distributions of the number of \pe\ measured with the F-factor method when illuminating the camera with flat-field events. Left: Histograms of \pe\ from HG and LG channels. Right: The number of \pe\ estimated in HG as function of the pixel position. Central pixels have a higher quantum efficiency by construction. }
    \label{fig:charge_calibration}
\end{figure}

Fig.~\ref{fig:charge_calibration} presents the \pe\ distributions obtained with flat-field events. The two gain channels\footnote{HG $\mathrm{\sim17}~\times$ LG} give equivalent results. The higher number of p.e. for the inner part of the camera reflects the PMT sorting based on QE, which divided the PMTs into higher and lower QE groups and placed the higher QE PMTs close to the center. The calibration coefficients, which permit conversion of ADC counts to a flat-fielded effective number of \pe, are given by
\begin{equation}
\mathrm{C} = \frac{1}{\mathrm{g}}  = \frac{\overline{\rm \pe}}{\overline{Q}-\overline{\rm ped}} ~,
\label{eq:FF_gain}
\end{equation}
where $\mathrm{g}$ is gain and $\overline{\rm \pe}$ is the camera median \pe

The real time calibration is obtained on a sample of 10,000 flat-field and 10,000 pedestal events, acquired at the beginning of the night with a rate of 1 kHz. 
The offline calibration is based on interleaved flat-field and pedestal events continuously acquired with a rate of 0.1 kHz, which allows an offline update of the calibration constants each 100 s.

\begin{figure}
    \centering
    \includegraphics[width=0.95\textwidth]{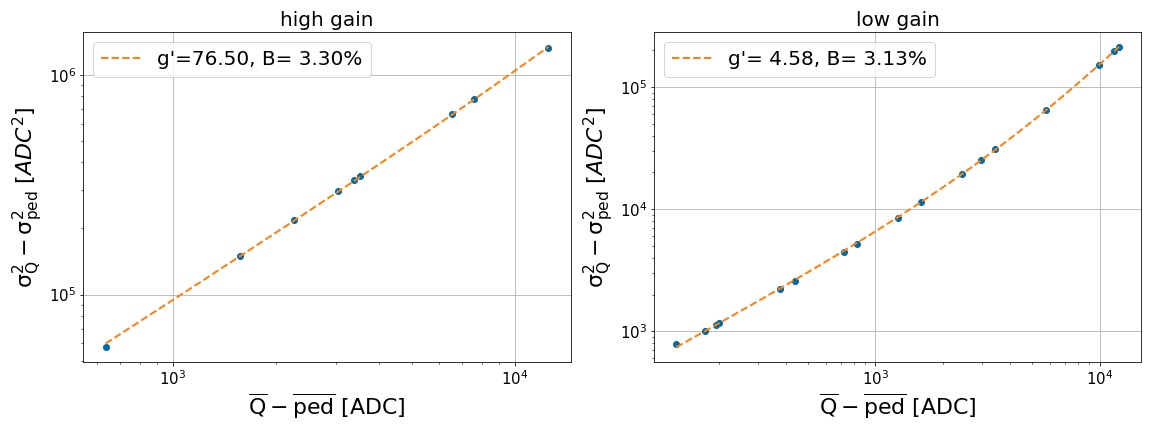}
    \caption{Signal variance as function of the charge (medians over the camera). The fit follows Eq.~\ref{eq:FF_gain_corrected}. Left: HG channel. Right: LG channel.}
    \label{fig:variance_fit}
\end{figure}

Finally, the calibration coefficients C are corrected by two global scaling factors. The first factor is necessary to take account of the different integration windows used for flat-field events (12 ns) and cosmic events (8 ns). The value is different for HG and LG due to the different pulse shape, and equal to 1.088 and 1.004, respectively. The second factor is due to the presence of an additional noise component that is proportional to the signal amplitude ($\sim \mathrm{B}\,Q$) and that is not taken into account in Eq.~\ref{eq:F-factor}. Then, in reality, the signal variance is described by
\begin{equation}
\sigma^2_Q-\sigma^2_{\rm ped} = \mathrm{g^\prime}\;\mathrm{F^2}  \; (\overline{Q}-\overline{\rm ped})+ \mathrm{B^2} \ (\overline{Q}-\overline{\rm ped})^2,
\label{eq:FF_gain_corrected}
\end{equation}
where $\mathrm{g^\prime}$ is the correct gain. The acquisition of flat-field events of different amplitudes, which can be obtained by setting proper filters in front of the laser emission point, allows us to estimate both the correct gain $\mathrm{g^\prime}$ and the $\mathrm{B}$ coefficient by fitting the signal variance as function of the charge, as shown in Fig.~\ref{fig:variance_fit}. A global scaling factor is then estimated by comparing the median fitted gain $\mathrm{g^\prime}$ with the median F-factor gain $\mathrm{g}$ and it is used to correct the calibration coefficients obtained, each night, with Eqs.~\ref{eq:F-factor} and \ref{eq:FF_gain}. The current factors are 1.08 and 1.09 for HG and LG, respectively. The additional noise term B is about 3\% for both channels. One of the major contributions to the systematic noise is effect of non-uniform time sampling of the waveform by the DRS4 chip \cite{2013NIMPA.723..109S}. It is estimated to be $\sim2.1\%$ in HG and $\sim2.7\%$ in LG. The values can be different between the two channels due to different pulse shapes. It is also estimated that there is $\sim0.7\%$ contribution from laser instabilities.
In the future, it is planned to update the calibration procedure in order to directly include the quadratic noise component, which can be fitted pixel per pixel, in the F-factor equations.
\section{Performance of signal reconstruction}

In order to examine performance of signal reconstruction by the current calibration chain, charge and time resolution for calibration pulses have been evaluated. A set of flat-field runs with different light intensities is analyzed to obtain performance at each signal amplitude. Charge and time of signals are reconstructed by pulse integration with 8 ns time window around the pulse peak. For absolute charge calibration, leak of charge out of the integration window is taken into account. The data were taken at a NSB level of $\sim400$ MHz, which is slightly higher than the dark NSB. Consistency with Monte Carlo (MC) simulation is also checked.

\subsection{Charge resolution}

The charge resolution is evaluated as the relative root-mean-square error (RMSE):
\begin{equation}
{\rm Rel.RMSE} = \frac{1}{N_{\rm sim}}\sqrt{{\rm Var(N_{\rm rec})} + b^2},
\label{eq:rel_rmse}
\end{equation}
where the $b$ term represents the charge reconstruction bias:
\begin{equation}
b = \langle N_{\rm rec} - N_{\rm sim} \rangle,
\end{equation}
and $N_{\rm sim}$ and $N_{\rm rec}$ are the number of simulated and reconstructed \pe, respectively. For the data, instead of the simulated number of \pe{}, absolute charge is estimated by charge extraction with a fixed window. The reason for using the fixed window is to avoid the bias which can be caused by searching for pulse peaks. The width of the fixed window is set to 29 ns so that the pulses are always fully integrated. Different charge extraction ratio from the pulses between standard 8 ns window around the pulse peak and the wide fixed window is taken into account for evaluating the bias. The absolute charge calibration is performed by the F-factor method at an intensity of $\sim80$ \pe/pixel as described in Section \ref{sec:charge_calib}.

\begin{figure}
    \centering
    \includegraphics[width=0.8\textwidth]{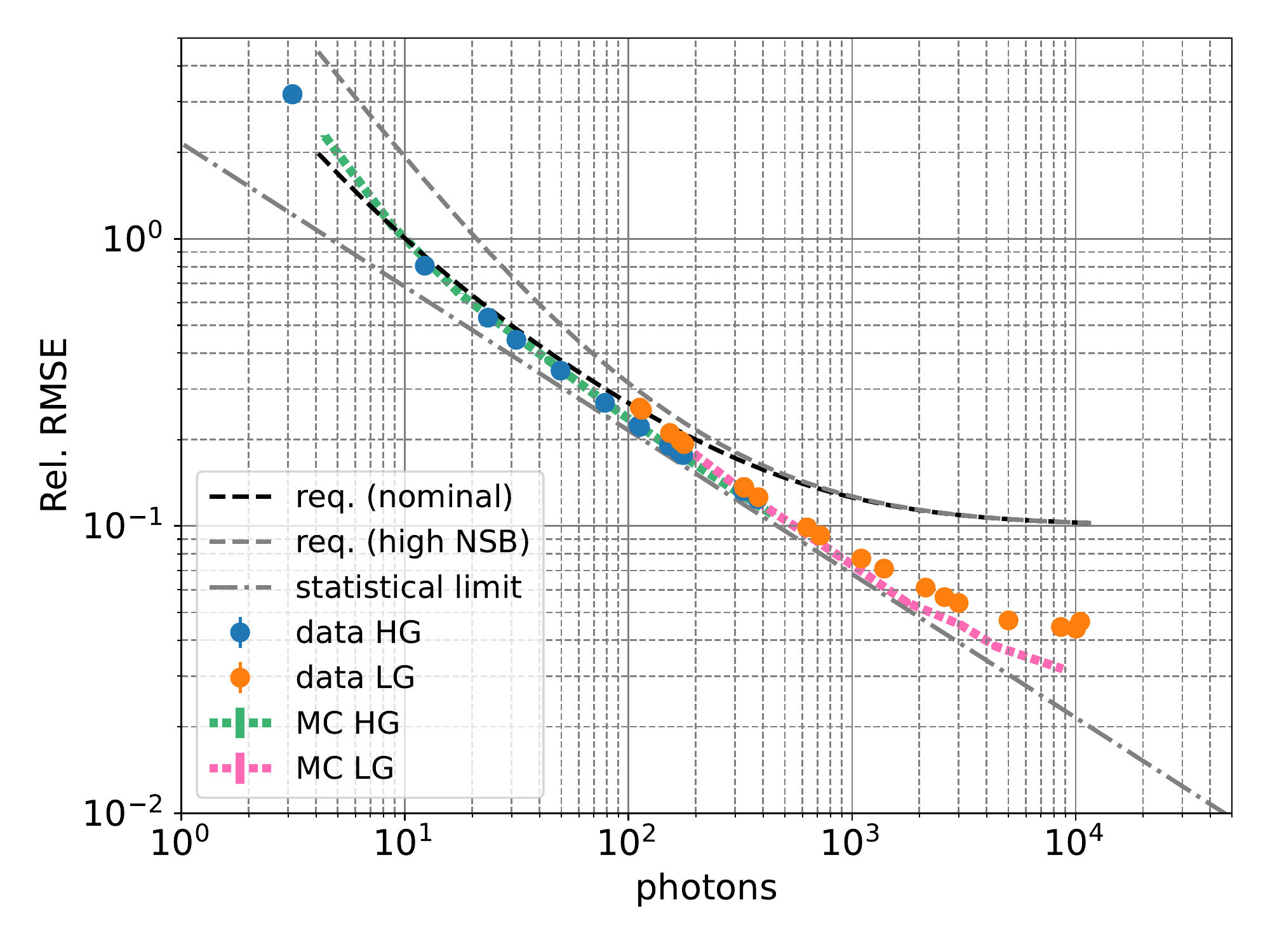}
    \caption{The relative RMSE of charge reconstruction evaluated from data and MC simulation. The requirements for two different NSB levels are shown by dashed lines. The dashed-dotted line is the statistical limit of the charge resolution. Photodetection efficiency is assumed to be 0.264.}
    \label{fig:rmse}
\end{figure}

Fig.~\ref{fig:rmse} shows the evaluated relative RMSE for both the data and MC simulation with comparison to the CTA requirements. The data points are average over the pixels. Efficiency of photon to \pe\ conversion is assumed to be 0.264 based on the latest evaluation of the LST-1 efficiencies. The results meet the requirement at most of the intensities. The slight violation at the lowest intensity can be due to the bias by charge extraction with pulse peak search. Note that when reconstructing Cherenkov photons, the position of the pulse integration window can be determined based on overall time evolution of extensive air showers, which should suppress the bias and improve charge resolution at low intensities. The worse resolution in the data than MC simulation above $\sim1,000$ photons can be explained by systematic uncertainties in charge reconstruction which are specific to the data (additional contributions to the $\mathrm{B}$ term, see Section \ref{sec:charge_calib}). The charge resolution of the data at high intensity region is apparently much better than the requirements because the requirements consider uncertainty in the absolute charge calibration while this is not taken into account for the data.

\subsection{Time resolution}
\begin{figure}
    \centering
    \includegraphics[width=0.54\textwidth]{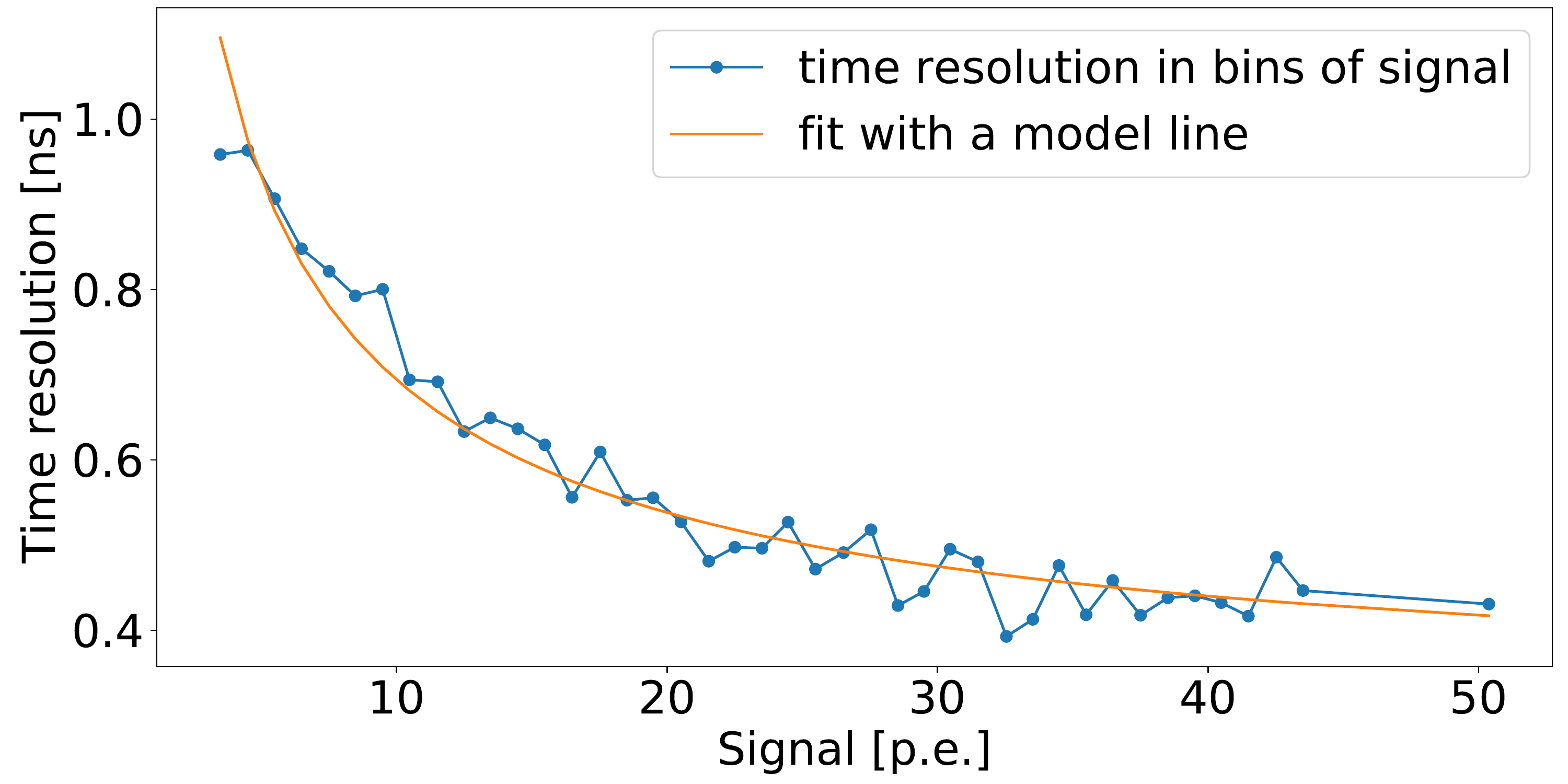}
    \includegraphics[width=0.44\textwidth]{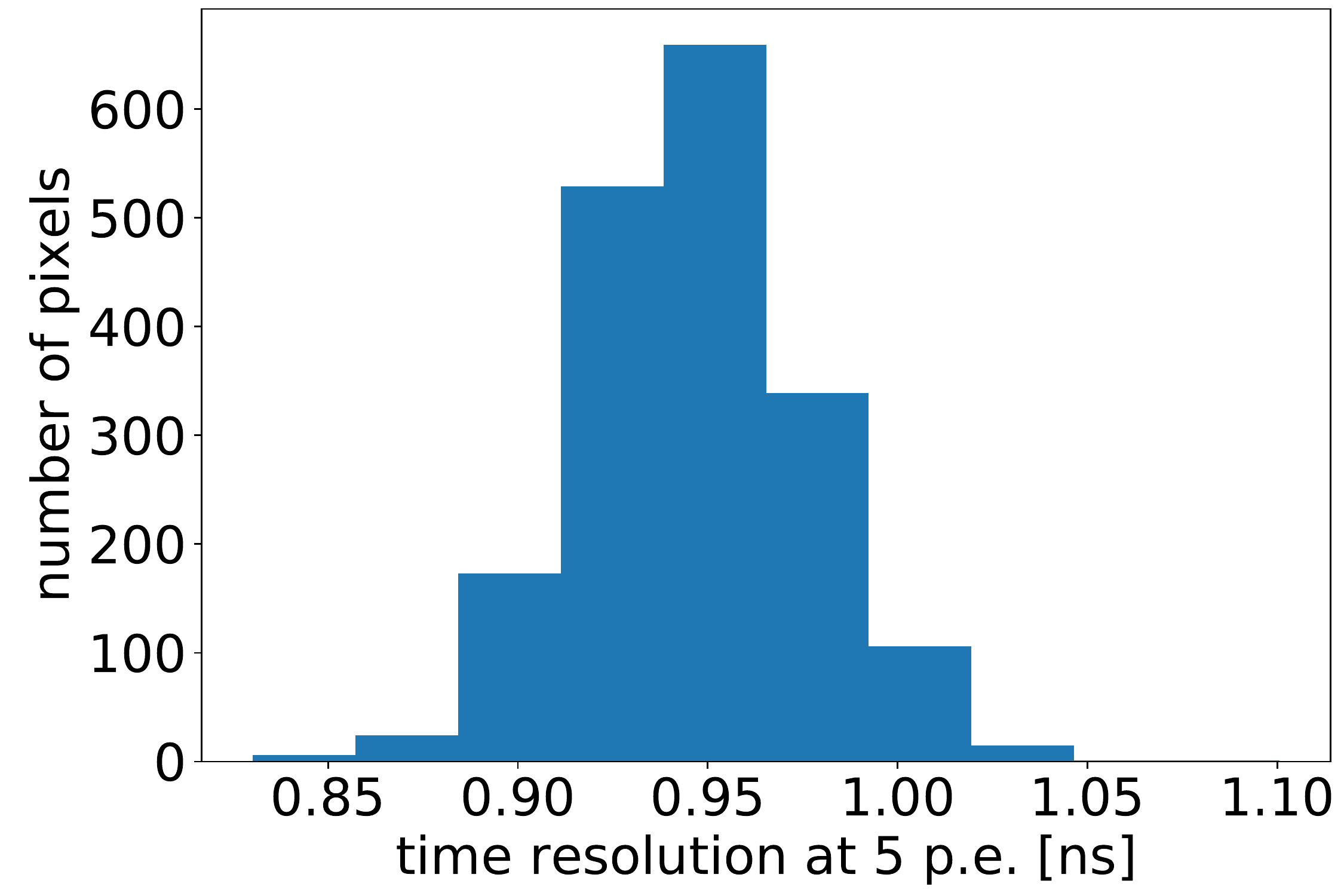}
    \caption{Left: the time resolution dependence on the signal strength (in \pe ) for an exemplary pixel.
    Right: distribution of the time resolution at 5\,\pe\ for all the pixels in LST-1.}
    \label{fig:timeres}
\end{figure}

For arrival time reconstruction of pulses we apply the DRS4, capacitor-wise, time correction.
To limit the extractor bias, the region of interest is limited to 26 ns. 
In order to avoid any global time jitter that would not affect the relative time resolution between individual pixels, e.g., jitter of the trigger timing, pulse time in each pixel is computed with respect to the average arrival time over pixels in each event. The average arrival time is calculated from all the pixels with a signal of at least 5\,\pe{}
The events in which there are less than 100 of such pixels are skipped from the analysis. 
We then bin the signals according to the reconstructed number of \pe\ and in each such bin, for each pixel, we compute the distribution of the arrival times, and fit it with a combination of a Gaussian function and a constant. 
The standard deviation of the Gaussian component is interpreted as the time resolution. 
Afterwards, for each pixel we fit the time resolution vs the number of \pe\ with a function allowing for Poissonian, linear and constant contributions (see \cite{2013NIMPA.723..109S} for details).
The example of such a fit is presented in the left panel of Fig.~\ref{fig:timeres}.

From such fits, we compute the time resolution for each pixel at 5\,\pe\ (see the right panel of Fig.~\ref{fig:timeres}). 
The differences from one pixel to another are not large, and a typical resolution is 0.95\,ns.
Applying the same procedure at MC simulation we obtain a slightly higher value of 1.2\,ns time resolution. 
The somewhat lower values in the data might be attributed to simplifications in MC simulations and it is under investigation.
Nevertheless, the difference is much smaller than the size of the extraction window and typical time cleaning threshold, hence its effect on the final analysis is expected to be small. 
It should be noted that the calibration pulses have a broader full width at half maximum than pulses from Cherenkov light, hence the time resolution for showers is expected to be slightly better than obtained here.

The camera of LST must fulfill a CTA requirement that the root mean square difference in the reconstructed signal arrival time for any two simultaneously illuminated pixels with amplitudes of 5 \pe{} must not exceed 2\,ns.  
This is roughly equivalent to the standard deviation of arrival times at each pixel to be below 1.3\,ns. Thus according to Fig.~\ref{fig:timeres}, the requirement is fulfilled for all the pixels. 
\section{Conclusions}

The calibration chain for LST-1 has been developed as part of the LST analysis pipeline. The calibration chain is applied to data and its performance has been evaluated. DRS4 pedestal corrections are confirmed to be working well. After applying all the corrections, pedestal noise level is $\sim0.2$~\pe\ in HG and $\sim3$~\pe\ in LG. Absolute charge calibration of the camera is performed by the F-factor method, and the result is compatible between the two gain channels. The global scaling factors are introduced in order to take account of effect of systematic noise in charge reconstruction and different charge extraction ratio depending on integration window and pulse shapes. In order to examine performance of signal reconstruction, charge and time resolution for the calibration pulses are evaluated. The charge resolution meets the requirement excluding the lowest intensities, where the charge reconstruction bias is significant. Note that the resolution at the lowest intensities will be improved for Cherenkov light by making use of overall time development of air showers. The time resolution obtained with the data is typically $\sim0.95$ ns at 5 \pe\ and fulfills the CTA requirement. We conclude that the calibration pipeline is ready for analysis of LST-1 observational data.

\section*{Acknowledgments}
We gratefully acknowledge financial support from the agencies and organizations listed here:
\url{www.cta-observatory.org/consortium\_acknowledgments}.
PG and JS are supported by the grant through the Polish National Research Centre No. 2019/34/E/ST9/00224.

\bibliographystyle{unsrt}
\bibliography{reference}
 


\clearpage
\section*{Full Authors List: \Coll\ The CTA LST Project}
%
%
\vspace{-3mm}
\scriptsize
\noindent
H. Abe$^{1}$,
A. Aguasca$^{2}$,
I. Agudo$^{3}$,
L. A. Antonelli$^{4}$,
C. Aramo$^{5}$,
T.  Armstrong$^{6}$,
M.  Artero$^{7}$,
K. Asano$^{1}$,
H. Ashkar$^{8}$,
P. Aubert$^{9}$
% ...

\vspace*{2mm}
\fontsize{8.35pt}{0cm}\selectfont
\noindent
$^{1}$Institute for Cosmic Ray Research, University of Tokyo.
$^{2}$Departament de FÃ­sica QuÃ ntica i AstrofÃ­sica, Institut de CiÃšncies del Cosmos, Universitat de Barcelona, IEEC-UB.
$^{3}$Instituto de AstrofÃ­sica de AndalucÃ­a-CSIC.
$^{4}$INAF - Osservatorio Astronomico di Roma.
$^{5}$INFN Sezione di Napoli.
$^{6}$Aix Marseille Univ, CNRS/IN2P3, CPPM.
$^{7}$Institut de Fisica d'Altes Energies (IFAE), The Barcelona Institute of Science and Technology.
$^{8}$IRFU, CEA, UniversitÃ© Paris-Saclay.
$^{9}$LAPP, Univ. Grenoble Alpes, Univ. Savoie Mont Blanc, CNRS-IN2P3, Annecy.
% ...

\end{document}